\documentclass[twocolumn,prb,showpacs,10pt]{revtex4}
\usepackage{psfrag}
\usepackage{graphicx}
\usepackage{amsmath}
\usepackage{amssymb}
\usepackage{subfig}

\usepackage{color}
\usepackage{multirow}
\begin{document}
\title{Destructive quantum interference in electron transport: A reconciliation of the molecular orbital and the atomic orbital perspective}

\author{Xin Zhao, Victor Geskin$^*$ and Robert Stadler}
\affiliation{\mbox{Institute for Theoretical Physics, TU Wien - Vienna University of Technology,}\\ Wiedner Hauptstrasse 8-10, A-1040 Vienna, Austria\\ 
Email: victor.geskin@mail.be, robert.stadler@tuwien.ac.at\\
$^*$Permanent address: 11, rue de la Limerie, B-7000 Mons, Belgium}

\date{\today}

\begin{abstract}
Destructive quantum interference (DQI) in single molecule electronics is a purely quantum mechanical effect and entirely defined by inherent properties of the molecule in the junction such as its structure and symmetry. This definition of DQI by molecular properties alone suggests its relation to other more general concepts in chemistry as well as the possibility of deriving simple models for its understanding and molecular device design. Recently, two such models have gained wide spread attention, where one was a graphical scheme based on visually inspecting the connectivity of carbon sites in conjugated $\pi$ systems in an atomic orbital (AO) basis and the other one put the emphasis on the amplitudes and signs of the frontier molecular orbitals (MOs). There have been discussions on the range of applicability for these schemes, but ultimately conclusions from topological molecular Hamiltonians should not depend on whether they are drawn from an AO or a MO representation, as long as all the orbitals are taken into account. In this article we clarify the relation between both models in terms of the zeroth order Green's function and compare their predictions for a variety of systems. From this comparison we conclude that for a correct description of DQI from a MO perspective it is necessary to include the contributions from all MOs rather than just those from the frontier orbitals. The cases where DQI effects can be successfully predicted within a frontier orbital approximation we show to be limited to alternant even-membered hydrocarbons, as a a direct consequence of the Coulson-Rushbrooke pairing theorem in quantum chemistry.
\end{abstract}
\maketitle

{\it .. you have a program, for God's sake use it, play with it, do a calculation on any small problem related to your problem. Let the calculations teach you. They are so easy! Shall we stop teasing one another about MO and VB?  

... quantum chemistry has given us two wonderful tools to reason about chemistry, and denying any one of them would impoverish our ability to reason.~\cite{header}}

\section{Introduction}

First experimental studies of electron transport through single molecules attached to metal contacts by using a scanning tunnelling microscope (STM) or mechanically controlled break junction techniques~\cite{chavy,tour,lohneysen,ruitenbeek}, also triggered considerable theoretical activity in this field since the beginning of the new millennium. The theoretical framework most widely used in this context is a Non-Equilibrium Green's function (NEGF) formalism~\cite{keldysh}, where coherent electron transmission according to Landauer's theory is assumed. 

The conductance of a molecular junction can then be described in dependence on the incoming electron's energy $E$ in terms of the transmission probability $T(E)$, which within NEGF is defined as
\begin{equation}\label{negf}
T(E)=Tr[G^{r}(E){\bf \Gamma}_{L}(E)G^{a}(E){\bf \Gamma}_{R}(E)],
\end{equation}
where the self energy matrices ${\bf \Gamma}_{L}$ and ${\bf \Gamma}_{R}$ contain the coupling of the molecule to the left and right electrodes and $G^{r}$ and $G^{a}$ are the retarded and advanced Green's functions (GF) of the (extended) molecule. 

The NEGF formalism has been implemented in a variety of codes, where in combination with density functional theory (DFT) it allows for a first principles treatment of the conductance of single molecule junctions~\cite{atk}$^{-}$\cite{kristian}. The usefulness of such calculations, however, still relies on simple models for interpreting them in terms of quantum chemical concepts such as atomic orbitals (AOs) or molecular orbitals (MOs) in order to achieve a qualitative understanding of the observed electron transport characteristics in the context of our general knowledge of the electronic properties of molecules or what is regarded as chemical intuition. 

In principle, electron transmission can be viewed as a particular manifestation of the more general phenomenon of electronic communication through a molecule, where a Green's function describes the propagation of a perturbation and is a measure of the connected paths made available by the  bonding pattern of the molecule~\cite{linderberg}. 

The zeroth Green's function $G^{mol}_{lr}$ can be used to rewrite the expression for $T(E)$ given in Eq.~\ref{negf} under the assumption that the central molecule is coupled to both metal contacts only via a single AO labeled $l$ and $r$ on the respective side, because then each matrix ${\bf \Gamma}$  contains only one non-zero element, $\Gamma_{ll}$ and $\Gamma_{rr}$, respectively and therefore a single term remains from the trace resulting in
\begin{equation}\label{trans}
T(E)=|G^{mol}_{lr}(E)|^2 \Gamma_{ll}(E) \Gamma_{rr}(E).
\end{equation}

By evaluating the relevant matrix elements of $G^{mol}_{lr}$, chemical understanding of the general properties of molecules can arguably be complemented by studying their transmission properties. For example, the low-bias conductance through benzene is orders of magnitude lower when it is contacted at positions which are meta (\textit{m}) with respect to each other when compared with ortho (\textit{o}) or para (\textit{p})~\cite{joachim1}$^{-}$\cite{mayor}. 

This result comes as no surprize for a chemist even without any prior exposure to the theory of molecular conductivity, who knows that the influence of a substituent in a benzene ring is "felt", in both electrophilic and nucleophilic substitution reactions, in o- and p-position to it, while m-positions "do not communicate". 

This knowledge might be referred to as chemical intuition but is actually based on rules stemming from resonance theory within the Valence Bond (VB) framework, where the relation of electronic communication to the topology of MOs is not self-evident but should be contained in $G^{mol}_{lr}$. In their classical 1947-1950 series of papers Coulson and co-workers attempted to put the one-electron Green's Function (without using the term at the time) at the heart of chemical theory~\cite{coulson1}$^{-}$\cite{pickup}. 

They demonstrated how starting from a H\"{u}ckel Hamiltonian in AO representation basic molecular characteristics such as MO energies, atomic charges, bond orders and response coefficients can be derived directly from the secular determinant without referring to explicit MO vectors, where the relation of this work to electron transport phenomena has been commented on very recently~\cite{stuyver1}$^{-}$\cite{diradical}.

Although it was correctly claimed by Datta~\cite{datta} amongst others that in a single molecule junction the conductance is defined not only by the central molecule but rather by the entire system including the metal contacts, the individual contributions of the components are separable in Eq.~\ref{negf}. Therefore, for the purpose of device design the molecular contribution can be optimized independently from the coupling to the metal contacts, a notion which has been recently confirmed in a joint theoretical and experimental work by Manrique et al.~\cite{lambert}. In this study it was shown that molecules and even their fragments contribute well defined and transferable factors to electron transport as a crucial observation for the investigation of destructive quantum interference (DQI) effects, a phenomenon which has been the topic of a tremendous number of recent articles, where for a rather complete bibliography we refer to Ref.~\cite{reuter}. 

Such DQI effects when occurring in the transmission close to the Fermi energy $E_F$ can be used for data storage~\cite{memory}, inducing thermoelectricity~\cite{fano} or the design of logic circuits~\cite{graphical1}. Simple models have been proposed for their analysis, which were derived from tight-binding (TB) or topological H\"{u}ckel theory and validated by DFT calculations: One of them which we refer to as "the graphical AO scheme" in the following has been derived specifically for the prediction of DQI and is based on a graphical analysis of the connectivity matrix of atomic orbitals (AOs)~\cite{graphical1}$^{-}$\cite{graphical5}, while the other interprets the efficiency of transmission in a broader sense in terms of the signs and amplitudes of molecular orbitals on the atomic sites directly connected to the electrodes~\cite{yoshizawa1}$^{-}$\cite{yoshizawa6}, and the analysis is sometimes limited to a "frontier orbital approximation" where only the highest occupied (HOMO) and the lowest unoccupied MO (LUMO) are taken into account.

The aim of this article is to reconcile the predictions from these two conceptually different approaches for an interpretation and analysis of the molecular Green's function. It is expected that by focusing on either an AO or a MO representation of the same quantum-mechanical problem one should obtain the same results. Their reconcilation is akin to that of the VB and MO theories in the earlier days of quantum chemistry by Van Vleck et al. in 1935~\cite{vanvleck}. 

But while VB and MO approaches become variationally equivalent for the ground state only in the limit of full configuration interaction, for electron transport within a single particle framework the representations of the molecular Green’s function in the AO and MO bases are already strictly equivalent on a single determinant level. From Eq.~\ref{trans} it can be seen that it is both necessary and sufficient to evaluate the purely molecular quantity $G^{mol}_{lr}$ for estimating whether the transmission will be finite or zero at any given energy $E$. 

The derivation of both "the graphical AO scheme" and the MO based scheme mentioned above start from this observation. Within a frontier orbital approximation, however, only the HOMO and LUMO are taken into accound instead of all MOs contained in $G^{mol}_{lr}$ and this approximation then limits the range of applicability of the MO based scheme to that of the Coulson-Rushbrooke pairing theorem~\cite{pairing,proof} as we explain in detail in the next section. 

If the contributions of all MOs and not only the frontier orbitals to $G^{mol}_{lr}$ in Eq.~\ref{trans} are correctly accounted for on a quantitative level, however, DQI can be analyzed from a MO perspective leading to equivalent results as the graphical AO scheme from Refs.~\cite{graphical1}$^{-}$\cite{graphical5} for all conjugated $\pi$ systems both alternant and non-alternant, with and without hetero-atoms and regardless of which subset of sites the contact atoms belong to, which is the main message of our article.

The paper is organized as follows: In the next section we shortly review the graphical AO scheme and highlight its relation to Eq.~\ref{trans}. Here we also explain on the basis of the pairing theorem that DQI effects entering $G^{mol}_{lr}$ can in general only be quantitatively described and understood in terms of the onsite energies of all MOs and their respective amplitudes at the contacted atomic sites. Furthermore, we clarify the connection of such a MO centered analysis scheme to Larsson's formula, which has been originally proposed for the definition of an effective coupling from the MO contributions to the transfer integral in a Marcus theory description of electron hopping~\cite{larsson}$^{-}$\cite{georg} but later on also used for the analysis of coherent electron transport in single molecule junctions~\cite{sautet1,no2bipy}.

In the following section we provide computational studies for a variety of test systems in order to substantiate our claim that it is possible to gain understanding of DQI effects in accurate terms for any conjugated $\pi$ system without the limitations of applicability facing the original frontier MO rules. 

For all the molecular systems in our article numerical calculations on a DFT level exist in the literature and most of them have also been studied experimentally. Since the focus of our work is on topological properties of $G^{mol}_{lr}$, for the calculations we present here topological H\"{u}ckel Hamiltonians are used in combination with NEGF. In this numerical chapter we also present the respective predictions from the graphical AO scheme for all systems as a reference and demonstrate their equivalence to the results obtained from an analysis of MO contributions, where the convergence with respect to the number of MOs included plays a prominent role. In the final chapter we provide a summary.

\section{Theoretical discussion of DQI prediction and analysis in AO and MO representations}

The zeroth Green's function $G^{mol}_{lr}$ in Eq.~\ref{trans} describes the propagation of a tunnelling charge between the atomic sites $l$ and $r$ mediated by all molecular orbitals (MOs) which in the weak coupling limit can be formulated as~\cite{graphical2}
\begin{equation}\label{zero}
G^{mol}_{lr}(E)=[(E \pm i \eta) {\bf I}-{\bf H}_{mol}]^{-1},
\end{equation}
where ${\bf H}_{mol}$ is  the molecular Hamiltonian, I a unity matrix and $i \eta$ an infinitesimal imaginary term introduced in order to avoid divergence of $G^{mol}_{lr}$ at the eigenvalues of  ${\bf H}_{mol}$.

\subsection{Graphical AO scheme}

Since $G^{mol}_{lr}$ is obtained from the inversion of the Hamiltonian matrix ${\bf H}_{mol}$, which is defined in an AO representation, one can analyze the properties of $G^{mol}_{lr}$ from the ratio of one of the minors of ${\bf H}_{mol}$ and its determinant~\cite{graphical1}$^{-}$\cite{graphical5} where $G^{mol}_{lr}$ is only equal to zero when the respective minor, as defined by the connection of the leads to two particular atomic sites $l$ and $r$ on the molecule, is also zero. 

In this way the graphical AO scheme for the prediction of DQI effects was derived, which has been formulated as the following set of rules: DQI occurs at $E=E_F$ if it is impossible to connect the two atomic sites $l$ and $r$ in a molecular topology, i.e. the only two sites with a direct coupling to the leads, by a continuous chain of paths, and at the same time fulfill the conditions (i) two sites can be connected by a path if they are nearest neighbors and (ii) at all atomic sites in the molecule other than $l$ and $r$, there is one incoming and one outgoing path. 

In other words, for a finite zero-bias conductance all AOs of the molecular topology have to be either traversed within a continuous chain of paths from $l$ to $r$ or be part of a closed loop in the topology, where the latter can be a double line due to the pairing of connected orbitals or a triangle or any larger loop~\cite{graphical1,graphical2,graphical4}. We will demonstrate in the following section how to apply these rules for any given molecular topology in praxis. In a later extension of this scheme it has been clarified that such defined paths can also cancel each other out in special cases and that therefore a sign has to be attributed to them~\cite{graphical5}. It has to be noted that these "paths" are just mathematical terms coming from forming the minor of ${\bf H}_{mol}$ and should not be interpreted as the physical path of an electron moving through the molecule. 

This graphical AO scheme has the advantage that it allows for the prediction of DQI without any numerical calculations being required simply by a visual assessment of the chemical structure of the central molecule in the junction. The scheme has been designed for molecules with a conjugated $\pi$ system, because it is only $\pi$ electrons which are taken into account in the topological H\"{u}ckel Hamiltonian it was derived from. In praxis this is not really a limitation, since potential functional molecules of interest are usually conjugated systems, where $\pi$-transmission is dominant. In order to allow for a simple analytical treatment, the derivation of the scheme also originally assumed sites with identical onsite energies and couplings to each other~\cite{graphical1,graphical2}. 

This assumption was later lifted in an attempt to generalize the method now also allowing for hetero atoms in the molecular structure but this came at the price of increased mathematical complexity~\cite{graphical3,graphical6}. Another assumption was that the only energy $E$, where $T(E)$ is of interest is the Fermi energy because it defines the zero-bias conductance and therefore the rules only apply at $E=E_F$. This latter assumption is rather delicate considering that in the model Hamiltonian the graphical rules were derived from the onsite energy of carbon sites were artificially set to $E_F$. Quite surprisingly, it was found in praxis that this rather crude approximation did not seem to limit the predictive qualities of the model even for cases where hetero atoms such as oxygen were involved in the molecular structures under investigation~\cite{graphical3} as long as the Fermi energy defined by the metal leads lies within the HOMO-LUMO gap of the molecule when energy levels are aligned~\cite{fermi}.

\subsection{Pairing theorem and frontier orbital approximation}

In order to gain a MO perspective of $G^{mol}_{lr}$ instead of an AO one, ${\bf H}_{mol}$ has to be looked at in its diagonalized form as ${\bf H}_{mol}={\bf C}{\boldsymbol \epsilon}_{MO}{\bf C}^{\dagger}$, where ${\boldsymbol \epsilon}_{MO}$ is the diagonal matrix of MO eigenenergies and ${\bf C}$ is the matrix of the coefficients for the expansion of all MOs as a linear combination of all AOs in the molecule. Inserting this definition of ${\bf H}_{mol}$ into Eq.~\ref{zero} gives
\begin{equation}\label{spectral}
G^{mol}_{lr}(E)=\sum^{N}_{m=1}\frac{C_{lm}C^{*}_{rm}}{E -\epsilon_m \pm i \eta},
\end{equation}
which is the spectral representation of $G^{mol}_{lr}$ in a H\"{u}ckel AO basis with $C_{l(r)m}$ the coefficient of the l(r)-th AO in the m-th MO in a sum that runs over all $N$ occupied and unoccupied MOs, which result from the coupling of the AOs defining the basis vectors for ${\bf H}_{mol}$. 

It should be stressed that Eq.~\ref{spectral} is exact for any Hamiltonian with an orthogonal AO basis and that this spectral representation of $G^{mol}_{lr}$ served as the starting point for the formulation of the molecular orbital rules for efficient transmission by Yoshizawa and co-workers~\cite{yoshizawa1}$^{-}$\cite{yoshizawa6}. 

For the special case of alternant hydrocarbons (AH), which are molecules with a conjugated $\pi$ system where carbon atoms can be divided into two subsets, "starred and unstarred", such that the atoms of one subset are bonded only to those from the other, the Coulson-Rushbrooke pairing theorem~\cite{pairing,proof} applies which states that (i) the $\pi$ electron energy levels are symmetrically distributed about the zero energy level (which is assumed to be $E_F$ in single molecule junctions) and (ii) that each occupied MO obtained from diagonalizing the corresponding Hamiltonian in an orthogonal AO basis with an energy $- \epsilon_{MO}$ has its mirror image in the unoccupied region with the energy $+\epsilon_{MO}$, which regarding its shape differs only in the sign of all AO coefficients of one subset. 

In the following we focus on molecules with an even number of MOs, which we can then group in Eq.~\ref{spectral} into pairs of the contributions from the MOs whose energies are linked by the symmetry relation it defines, i.e (H,L), (H-1,L+1),...,(H-(N/2-1),(L+(N/2-1)) with H=HOMO, L=LUMO and $-\epsilon_{H-k}=\epsilon_{L+k}=\epsilon_k$. We can then redefine $G^{mol}_{lr}(E_F)$ as the sum of these pairs, which in the following we will refer to as Coulson-Rushbrooke or CR pairs:
\begin{equation}\label{coulson}
G^{mol}_{lr}(E_F)=\sum^{N/2-1}_{k=0}\frac{C_{l,(H-k)}C^{*}_{r,(H-k)}-C_{l,(L+k)}C^{*}_{r,(L+k)}}{\epsilon_k}.
\end{equation}

The pairing theorem now predicts for AOs $l$ and $r$ on carbon atoms of the same subset that $C_{l,(H-k)}C^{*}_{r,(H-k)}=C_{l,(L+k)}C^{*}_{r,(L+k)}$ because either $C_{l,(H-k)}=-C_{l,(L+k)}$ and simultaneously $C^{*}_{r,(H-k)}=-C^{*}_{r,(L+k)}$, or $C_{l,(H-k)}=C_{l,(L+k)}$ and simultaneously $C^{*}_{r,(H-k)}=C^{*}_{r,(L+k)}$, as all the coefficients in only one subset change their sign when comparing an occupied with its mirrored unoccupied level. Therefore, the terms in every CR pair of Eq.~\ref{coulson} cancel exactly at $E_F$ for this case and DQI occurs as a result as has also been observed in Refs.~\cite{mikkel1,mikkel2}.

If on the other hand the contact AOs $l$ and $r$ belong to carbon atoms from different subsets, then  $C_{l,(H-k)}C^{*}_{r,(H-k)}=-C_{l,(L+k)}C^{*}_{r,(L+k)}$ because either $C_{l,(H-k)}=-C_{l,(L+k)}$ and $C^{*}_{r,(H-k)}=C^{*}_{r,(L+k)}$, or $C_{l,(H-k)}=C_{l,(L+k)}$ and $C^{*}_{r,(H-k)}=-C^{*}_{r,(L+k)}$. For this case, the contributions coming from the two individual parts of each CR pair of MOs (H-k, L+k) including the HOMO and the LUMO always add up constructively at $E_F$ in Eq.~\ref{coulson}. 

Although any individual CR pair contribution is therefore nonvanishing, it is important to stress that destructive interference is still possible between CR pairs, as each of them can contribute either a positive or a negative term to $G^{mol}_{lr}$. The pairing theorem, however, does not provide the means for an assessment of prediction of such inter-pair interference. 	

The general conclusion from the pairing theorem is therefore that DQI will always occur for the electron transport through junctions containing alternant hydrocarbons when carbon atoms of the same subset are contacted, which is already sufficient to account for the low conductance of a variety of systems such as polyenes with contact atoms of the same parity, meta-contacted benzene or generic cross-conjugated molecules, where these cases can readily be identified from their chemical structure without any deeper analysis of the shapes and signs of their frontier MOs. 

On the other hand, for alternant hydrocarbons contacted on carbons belonging to different subsets, i.e. where one contact atom is starred and the other one unstarred or for non-alternant hydrocarbons or for conjugated $\pi$ systems containing hetero atoms, the pairing theorem can neither predict nor rule out DQI. In the literature these two cases are sometimes distinguished in terms of "easy zeros" (the same subset contacted) and "hard zeros" (different subsets contacted)~\cite{polyenes} or linked to the occurence of an odd or even number of zeroes in $T(E)$~\cite{mikkel1,mikkel2}. But for the purpose of our article the important distinction is that for even-membered alternant hydrocarbons contacted at sites of the same subset DQI will always occur, while for all other cases DQI cannot be predicted without numerical calculations from a MO perspective.

We note that our discussion above only refers to alternant hydrocarbons with an even number of MOs and therefore also an even number of carbon sites. This is the general case for stable alternant hydrocarbons. When the total number of MOs is odd, which implies the existence of a non-bonding MO at the Fermi energy with non-vanishing contributions from only one subset follows from the pairing theorem, which then allows for  a conduction peak instead of a DQI induced minimum at $E_F$ when the contacted atoms belong to the subset contributing to this non-bonding MO.

We now turn our attention to the molecular orbital rules derived by Yoshizawa and co-workers~\cite{yoshizawa1}$^{-}$\cite{yoshizawa6}, where the starting point was also the spectral representation of $G^{mol}_{lr}$ given in Eq.~\ref{spectral}. These rules are amongst the earliest formulated providing a link between the complex phenomenon of DQI in electron transmission and the standard output of quantum chemical calculations, in this case the sign of the amplitudes of MOs. Within a frontier orbital approximation they also become particularly simple to apply because then the entire sum in Eq.~\ref{coulson} is dominated by only one CR pair, namely the contribution to $G^{mol}_{lr}(E_F)$ coming from the HOMO and the LUMO, and then the remaining pairs can all be neglected because their large energetic distance $\epsilon_k$ to $E_F$ results in large denominators in the respective terms, thereby making them numerically negligible. 

From this assumption, it can be concluded that the transport through a single molecule would be effective, i.e. DQI would be absent, when on the two contact atoms to the two leads i) the sign of the product of the MO expansion coefficients in the HOMO ($C_{l,H}C^{*}_{r,H}$) is different from that in the LUMO ($C_{l,L}C^{*}_{r,L}$) and ii) all four involved amplitudes $C_{l,H}$, $C^{*}_{r,H}$, $C_{l,L}$ and $C^{*}_{r,L}$ are of significant magnitude. If these conditions are not fulfilled, then "inefficient" transmission due to at least a partial cancellation of the contributions from the HOMO and LUMO was predicted which was not formulated as necessarily the zero transmission which is typical for DQI in a rigid sense. 

Such a frontier orbital approximation, however, only delivers correct results for the prediction of DQI whre the CR pairing theorem~\cite{pairing,proof} is applicable. If the atoms contacted by the two electrodes belong to the same subset (either starred or unstarred) of carbon atoms in an even-membered AH, the cancellation of the contributions from the HOMO and the LUMO to $G^{mol}_{lr}(E_F)$ is a reliable indicator of DQI not necessarily because they are dominant, but because it represents the cancellation of also the contributions within all other CR pairs entering Eq.~\ref{coulson}. This is the reason why DQI can be understood in this case in terms of the frontier orbitals alone.

For all other cases all MOs in the system need to be considered. If an alternant hydrocarbon is contacted at atomic sites belonging to different subsets, i.e one being starred and one being unstarred according to the CR framework, then although contributions from the HOMO and LUMO can only interfere constructively the tails related to lower lying occupied and higher lying unoccupied MOs might still cancel out with those of the frontier orbitals at $E_F$ and cause DQI. For non-alternant hydrocarbons and organic molecules containing hetero atoms, it turns out to be equally insufficient to limit the analysis to just one or even two CR pairs of MO contributions. In the next section we will provide a range of numerical examples justifying this assertion.

\subsection{Larsson's formula for a MO based analysis}

A somewhat simplified form of Eq.~\ref{spectral} has been known for decades as Larsson's formula in a different but related context, where it was used for the definition of the transfer integral mediated by a selected set of MOs in a Marcus theory description of electron hopping~\cite{larsson}$^{-}$\cite{georg}. More recently, it has been realized~\cite{sautet1,no2bipy} that the same formula can be also employed to define an approximation for $T(E)$ in coherent eletron tunnelling as $T(E)\sim\Gamma^2(E)$ with $\Gamma(E)$ being an energy dependent effective coupling containing the contributions from all MOs of a molecular bridge and defined by
\begin{equation}\label{gamma}
\Gamma(E)=\sum^{N}_{m=1}\frac{\alpha_{m}\beta_{m}}{E-\epsilon_{m}}.
\end{equation}

Here $\varepsilon_{m}$, $\alpha_{m}$ and $\beta_{m}$ are the eigenenergy, and the respective couplings to the left and right contact of the $m$th MO and $E$ is the kinetic energy of a transferred electron. It is easy to see by direct comparison that the effective coupling $\Gamma(E)$ in Eq.~\ref{gamma} is very much related to the zeroth Green's function in Eq.~\ref{spectral}. There are only two differences between the two equations. First, the MO amplitudes $C_{l(r)m}$ have been replaced by couplings $\alpha_{m}$ and $\beta_{m}$, which describe the overlap of each MO with a respective contact AO on the two leads. Since in the NEGF-TB description we employ for our numerical studies in the following section the contact AO on the leads is always the same orbital, this difference amounts to just the same constant factor for all the MO terms of the sum in Eq.~\ref{gamma} and is therefore irrelevant for our study where we just set this value to 1. 

The second difference between the two expressions, namely the dropping of the infinitesimal term $i \eta$ just means that $\Gamma(E)$ diverges at the eigenergies of all MOs. In principle this deficiency can be repaired by introducing an energy dependent normalisation factor as has been derived from the more general theory in Ref.~\cite{sautet2} by Sautet et al.~\cite{sautet1} under the very limiting condition of $\alpha_{m}>>\beta_{m}$, which is relevant when the focus is on the analysis of STM measurements. Since the qualitative behaviour of $\Gamma^2(E)$ does not deviate from that of $T(E)$ obtained from NEGF-TB in all systems investigated in this article, we avoide such a normalisation factor as an unnecessary complication. In the following section we just truncate $\Gamma^2(E)$ as obtained from Eq.~\ref{gamma} at the poles and scale it with the arbitrary constant of $10^{-2}$ for the purpose of its graphical presentation in the related figures.

While the poles of $G^{mol}_{lr}(E_F)$ in Eq.~\ref{spectral} or $\Gamma(E)$ in Eq.~\ref{gamma} define the peaks in $T(E)$ when these quantities are squared and each of these peaks can be identified with the electron transmission through one individual MO in the absence of degeneracies, the non-resonant transmission for energies between the peaks contains contributions from all MOs $m$ with the respective couplings $\alpha_{m}$ and $\beta_{m}$ and these contributions can interfere constructively or destructively in dependence on the energy $E$. 

In the sums of Eqns.~\ref{spectral} and ~\ref{gamma} the sign of the contribution from any MO is determined by i) the numerator of its corresponding term in the summation as defined by the product of amplitudes at the contact site or product of couplings to the metal leads and ii) its denominator which depends on whether the energy $E$ for which $T(E)$ is evaluated lies below or above the onsite energy of the MO in question. For AHs contacted via carbon atoms belonging to the same subset, DQI at the center of the HOMO-LUMO gap can be directly concluded from the CR pairing theorem by using Eq.~\ref{coulson}. 

For all other cases, neither the pairing theorem applies nor can reliable predictions on DQI be obtained within a frontier orbital approximation, because the occurrence or absence of DQI for any given value of $E$ seems to depend on a fine balance of cumulative contributions with different signs from all MOs, where a quantitative description of the decay of their respective tails is crucial as we will demonstrate in the next section.

\section{Numerical case studies}

\subsection{Computational details}
All the case studies in this article are based on simple models both in AO and MO representations as derived from the topological properties of the molecular Hamiltonian ${\bf H}_{mol}$ in a tight binding approximation with all onsite energies for the carbon AOs contributing to the $\pi$ system set to zero, i.e. to the origin of energy assumed to represent $E_F$, and the couplings between them to the resonance integral $\beta$ from H\"{u}ckel theory which also defines the unit of the energy axis. 

As an appropriate numerical benchmark for our conclusions we therefore present NEGF-TB calculations which have been conducted within the Atomic Simulation Environment (ASE)~\cite{ase1,ase2} with a coupling of $\beta$ between the molecular topology and the semi infinite carbon chains used for the electrodes. 

Sulfur atoms have been given an onsite energy of 1.11 $\beta$ and C-S bonds a coupling value of 0.69 $\beta$~\cite{hulis} in order to account for the effect of the hetero atom. Since the molecules we investigate here all have been chosen due to the recent interest they attracted, we also refer to the relevant literature for each system in order to show that our conclusions harmonize with the results from more realistic NEGF-DFT calculations or experimental conductance measurements.

\begin{figure*}
    \includegraphics[width=0.8\linewidth]{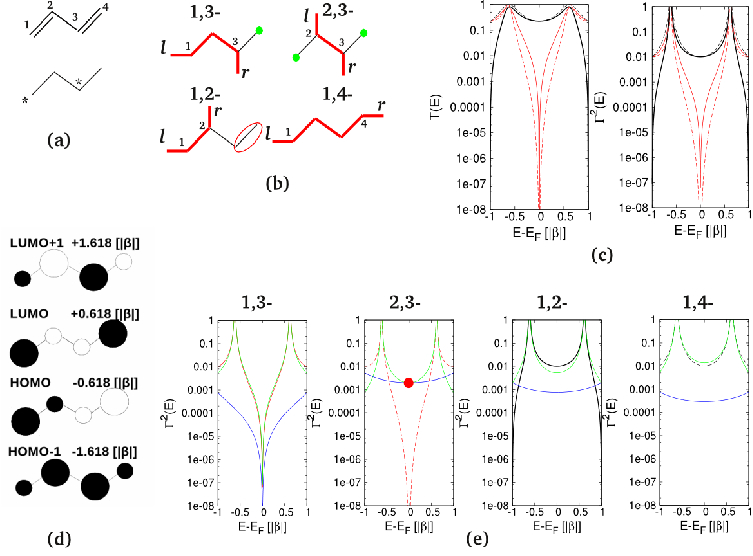}
    \caption{Butadiene contacted to two electrodes with all possible combinations of contact sites: a) chemical structure and "starring scheme", b) application of the graphical AO scheme (for details we refer to the caption of Fig.~\ref{fig4}), c) T(E) (in units of the conductance quantum $G_0$) from NEGF-TB compared with $\Gamma^2(E)$ (in arbitary units) from Eq.~\ref{gamma} with red lines where DQI occurs (solid for 1,3- and dashed for 2,3-) and black lines where it does not (solid for 1,2- and dashed for 1,4-), d) amplitudes (indicated as the size of spheres at each atomic size with black and white fillings distingusihing between a positive or negative sign of the wavefunction, respectively) and energies of all $\pi$-MOs, e) individual contributions from the (HOMO+LUMO) (green line) and (HOMO-1+LUMO+1) (blue line) to $\Gamma^2(E)$ (where the crossing point of the two curves is marked with a red dot for the 2,3-connection) which is contrasted with the total $\Gamma^2(E)$ from c) in the respective color code used there.}
    \label{fig1}
\end{figure*}

\subsection{Butadiene as the simplest illustrative example}

As a first example for our arguments in the last section we consider butadiene contacted at different sites (Fig.~\ref{fig1}). Since this molecule with a conjugated $\pi$ system is an even-membered AH, its even- and odd-numbered atoms belong to different starred/unstarred subsets and therefore DQI can be predicted for the (1,3-) connection within a frontier orbital approximation in agreement with the pairing theorem as outlined in section II B because carbon sites belonging to the same subset are contacted (Fig.~\ref{fig1}a). For all other possible connections, sites belonging to different subsets are contacted, and therefore constructive interference of the contributions from the HOMO and LUMO alone is found according to the pairing theorem. 

The graphical AO scheme (Fig.~\ref{fig1}b) as well as NEGF-TB calculations for $T(E)$ and their estimates as $\Gamma^2(E)$ from Eq.~\ref{gamma} after a diagonalisation of ${\bf H}_{mol}$ where for both all four MOs have been properly accounted for (Fig.\ref{fig1}c) find DQI not only for the 1,3- but also for the 2,3-connection. Within the graphical AO scheme DQI is predicted if it is not possible to form a continuous line between the two contact sites and have all AOs which are not on this line grouped up in pairs or as part of a closed loop. The single sites which are not crossed or grouped up are marked by green dots for the sake of clarity, which also allows for the simple correspondence that DQI occurs where green dots are unavoidable.

The MO based scheme within a frontier orbital approximation on the other side also correctly predicts DQI for 1,3-positioning of the contacts, but not for the 2,3-connection. While the amplitudes of both the HOMO and LUMO are low on sites $2$ and $3$ (Fig.~\ref{fig1}d), this justifies a reduced conductance but not a cancelling out to zero which is characteristic for DQI and found for the 2,3-connection for $T(E)$. 

In Fig.~\ref{fig1}e we plot $\Gamma^2(E)$ when only the contributions from the HOMO and the LUMO enter the expression for $\Gamma(E)$ in Eq.~\ref{gamma} (green lines). There it can be seen that indeed zero transmission is found also considering only the two frontier orbitals for the 1,3-connection in agreement with the predictions from the pairing theorem, while the 2,3-case shows a reduced but finite conductance when compared to 1,2- and 1,4-positions of the contacts. 

We also plot $\Gamma^2(E)$ from the contributions of the HOMO-1 and LUMO+1 alone (blue lines) and find that only for the (2,3-) connection they cross those of the frontier orbitals at $E_F$. Since the corresponding sum of terms entering Eq.~\ref{gamma} for the two pairs have different signs at their crossing point, they cancel each other out and lead to zero transmission. This is probably the simplest example to contrast a case of DQI, which can be predicted within a frontier orbital approximation with one that is beyond its range of applicability.

\begin{figure}
    \includegraphics[width=\linewidth]{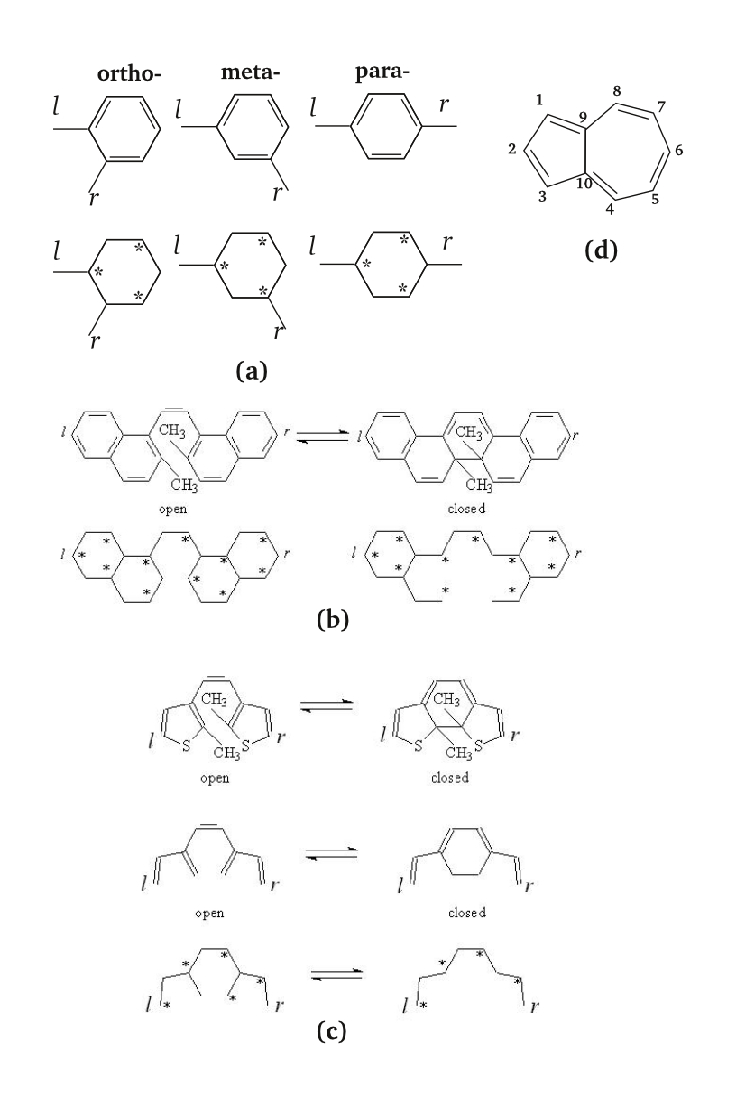}
    \caption{Chemical structures and corresponding TB next neighbor connectivity with the positions of the contacts to the electrodes marked as $l$ and $r$ and with "stars" related to the pairing theorem and involving only the $\pi$ electrons for a) benzene connected in ortho- (\textit{o}-), meta- (\textit{m}-) and para (\textit{p}-) positions, b) dinaphthylethene (DNE) and c) dithienylethene (DTE) in their open and closed forms, respectively, where for the latter also the alternant hydrocarbon analogs obtained by removing the S sites are shown. In d) the chemical structure of azulene is shown and the positions for connecting metal contacts are numbered for making reference to them in the text.}
    \label{fig2}
\end{figure}

\subsection{Other representative molecular structures}

In Figure~\ref{fig2} we show the other molecular systems which we investigate as case studies in this article and also the corresponding TB next neighbor connectivity, which provides the basis for all our NEGF-TB computations as well as the application of the pairing theorem and the graphical AO scheme in the following. We note that only unsaturated carbon atoms are part of the $\pi$ system and that it is only those which need to be considered in a TB framework. 

For benzene (Fig.~\ref{fig2}a), it is established knowledge~\cite{joachim1}$^{-}$\cite{mayor} that the conductance is finite for an \textit{o}- or \textit{p}-connected pair of contacts but DQI occurs at $E_F$ for a \textit{m}-connection, which is also consistent with the chemical understanding of communication through an aromatic ring. In Figure~\ref{fig2}a we illustrate that these findings can be also understood in the context of the pairing theorem (Fig.~\ref{fig2}b) because only for the m-connection two "starred" carbon sites are contacted in this example of an even-membered AH meaning that each CR pair will provide a contribution of exactly zero in Eq.~\ref{coulson}. 

Another type of systems where DQI plays an important role are molecular switches~\cite{switches} based on "conducting" and "insulating" isomers that can be transformed into each other in a highly reversible photochemical reaction. We will consider here one family of such switches, namely diarylethenes~\cite{irie}, and in particular the homocyclic dinaphthylethene (DNE) (Fig.~\ref{fig2}b) and the heterocyclic dithienylethene (DTE) (Fig.~\ref{fig2}c). For both systems the closed isomer is much better conducting than the open isomer in a molecular junction which has been demonstrated experimentally~\cite{switches,irie,irie1,launay,dulic} and confirmed theoretically~\cite{yoshizawa2,victor}, where the formation (or breaking) of a single bond distinguishes one from the other in structural terms. 

NEGF-DFT calculations can be found e.g. in Ref.~\cite{yoshizawa2} and Ref.~\cite{victor} for DNE and DTE, respectively. The molecular orbital rules by Yoshizawa and co-workers have also been applied for both systems, where although their application in a narrow sense would have suggested constructive interference for the "closed" (conducting) and "open" (insulating) form for DNE, the differences in conductance between the two forms found with NEGF-DFT has been attributed to the larger orbital amplitudes for the "closed" form~\cite{yoshizawa2}. As for the butadiene example we discussed above, this argument explains quantitative differences in the conductance but not the qualitative difference defined by the occurrence or absence of an interference minimum. 

For DTE on the other side, the contributions from the HOMO-1 and LUMO+1 had to be added to those from the frontier orbitals in order to reach better agreement with experimental findings~\cite{angewandte1}. As can be seen from their respective TB next neighbor connectivity in Figs.~\ref{fig2}b and c the electrodes are attached to carbon atoms which belong to different subsets of the starring scheme for the open forms of both DNE and DTE. As a consequence of the MO symmetry properties following from the pairing theorem therefore constructive QI has to be found within all CR pairs of MOs defined by Eq.~\ref{coulson} including the HOMO and the LUMO, and hence DQI can only occur due to cancellation of terms between different CR pairs which cannot be assessed by using a frontier orbital approximation, and this is the reason why we included these systems in the present study. 

DTE is heteorocyclic, which means that we need to include an onsite energy for the sulfur atoms differing from those of the carbon sites and also a value for the C-S coupling in the NEGF-TB calculations in the following as specified in the computational details. In order to have a reference point for the application of the graphical AO scheme also for DTE, we also introduce a structure with the S atoms of DTE deleted in Fig.~\ref{fig2}c, which transforms it into an even-membered AH, where in the next section we will compare the transmission functions of both DTE forms with and without sulfur sites.

As the final system for our study we chose azulene (Fig.~\ref{fig2}d), which is a non-alternant hydrocarbon and therefore it is not possible to divide its carbon sites into "starred" and "unstarred" subsets in relation to the pairing theorem or derive any conclusions regarding destructive or constructive QI within a frontier orbital approximation. This system is also of interest because it has been wrongly claimed in a joint experimental and theoretical study that for azulene the graphical AO scheme also fails in its predictions at least when the electrodes contact the sites $1$ and $3$ in Fig.~\ref{fig2}d~\cite{azulene}. 

This claim has later been refuted~\cite{graphical4}, where it was shown that the predictions of the graphical AO scheme were also correct for azulene with $1$,$3$-contacts when closed loops of AOs and not only pairs of them are considered which always has been a central aspect of the scheme~\cite{graphical1,graphical2}. NEGF-DFT calculations for azulene containing compounds with different electrode contact sites including the $1$,$3$- and $5$,$7$-cases can be e.g. found in Ref.~\cite{azulene} and Ref.~\cite{angewandte2}.

\begin{figure*}
    \includegraphics[width=0.7\linewidth]{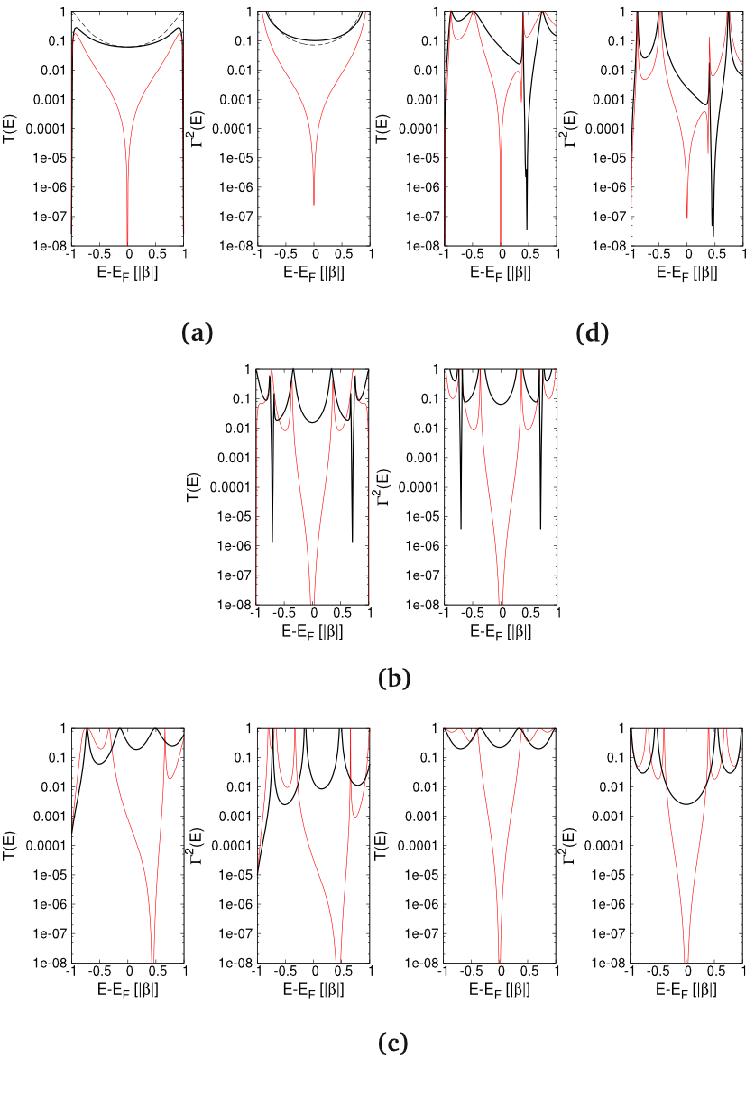}
    \caption{T(E) from NEGF-TB calculations (in units of $G_0$) is contrasted with $\Gamma^2(E)$ as obtained from Eq.~\ref{gamma} (in arbitrary units) for a) benzene connected in \textit{o}- (solid black), \textit{m}- (solid red) and \textit{p}- (dashed black) positions, b) DNE and c) DTE in their open (solid red) and closed (solid black) forms, respectively, where for the latter also curves for the AH analog of DTE with S atoms removed are shown in the two panels at the right side and d) azulene contacted in $1$,$3$ (solid black) and $5$,$7$ (solid red) positions.}
    \label{fig3}
\end{figure*}

\begin{figure}
    \includegraphics[width=\linewidth]{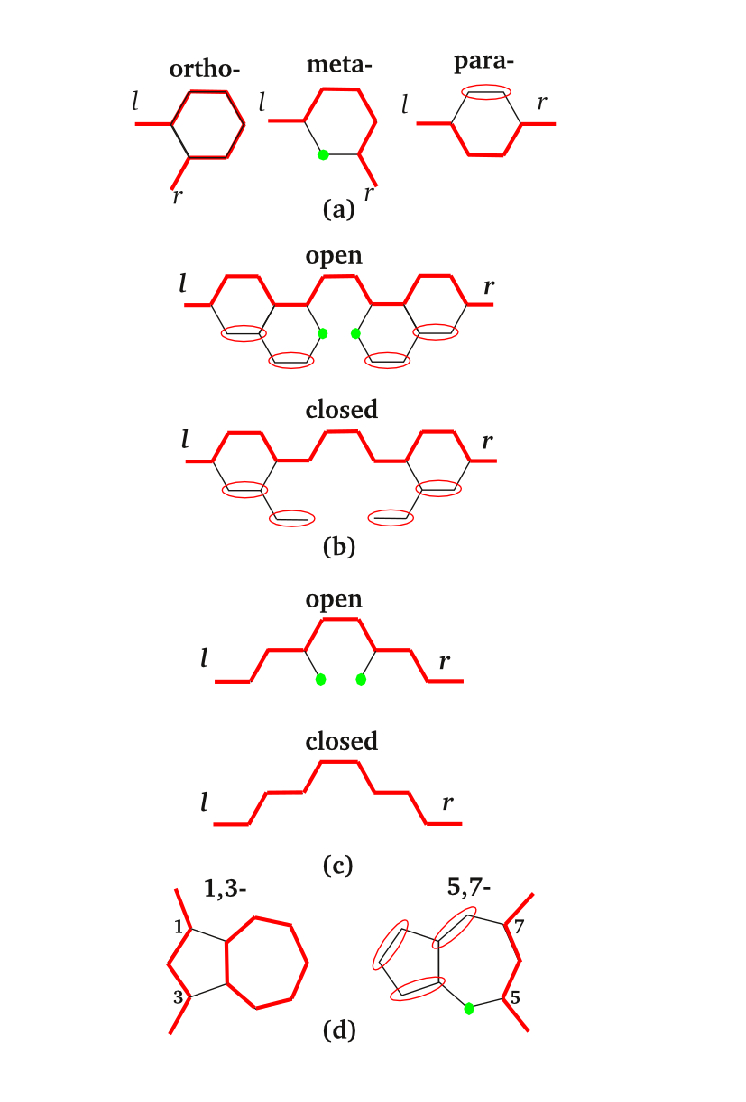}
    \caption{Application of the graphical AO scheme~\cite{graphical1,graphical2} for a) benzene connected in \textit{o}-, \textit{m}- and \textit{p}-positions, b) DNE and c) DTE (with S atoms removed) in their open and closed forms, respectively, and d) azulene contacted in $1$,$3$- and $5$,$7$-positions. DQI is predicted if it is not possible to form a continuous line between the two contact sites and have all AOs which are not on this line grouped up in pairs or as part of a closed loop. The single sites which are not crossed or grouped up are marked by green dots for the sake of clarity.}
    \label{fig4}
\end{figure}

\subsection{DQI predictions from NEGF-TB, Larsson's formula and the graphical AO scheme}

In Fig.~\ref{fig3}, $T(E)$ as calculated from NEGF-TB for all systems introduced in Fig.~\ref{fig2} is shown in the left panels for each label and compared with $\Gamma^2(E)$ in the right panels, which was obtained from Eq.~\ref{gamma} with MO onsite energies and amplitudes resulting from a diagonalisation of ${\bf H}_{mol}$ in the same TB setup, where for both quantities the curves for the systems exhibiting DQI at $E_F$ are shown in red and the others in black. 

Apart from the units, which are in multiples of the conductance quantum $G_0$ for $T(E)$ and chosen arbitrarily for $\Gamma^2(E)$ the agreement in all cases is excellent, which fully justifies to investigate the absence or occurrence of DQI solely in terms of the contributions entering Eq.~\ref{gamma}. 

From both $T(E)$ and $\Gamma^2(E)$ the m-connection is correctly identified as the only one with DQI for benzene (Fig.~\ref{fig3}a) and zero conductance found only for the open form of DNE (Fig.~\ref{fig3}b), a result which needs the inclusion of all MOs and not just the frontier orbitals~\cite{yoshizawa2} as we will further argue below. 

Also for DTE (Fig.~\ref{fig3}c) only the open form exhibits a transmission zero at $E_F$ for the analog alternant hydrocarbons (right two panels), which is shifted to higher energies if the S atoms are included in the calculations (left two panels) but still lowers the conductance at the Fermi energy in its vicinity quite substantially even then. 

For azulene (Fig.~\ref{fig3}d) there are QI minima across the energy spectrum for the two investigated junctions which differ in their respective contact sites. But while contacts in the $5$,$7$-positions (red lines) result in zero conductance, the minima for the $1$,$3$-connected system are not only lying above $E_F$ but are also so narrow that they do not seem to have an impact on $T(E_F)$. 

We note that all these features we summarized here are in good qualitative agreement with the respective NEGF-DFT calculations in the literature we referred to in the last section when introducing the respective molecular structures above.

In Fig.~\ref{fig4} we demonstrate the application of the graphical AO scheme~\cite{graphical1,graphical2} for all systems in Fig.~\ref{fig2} without hetero atoms and its predictions for DQI identify the cases with $T(E_F)=0$ in the calculations shown in Fig.~\ref{fig3} without a single failure, regardless of whether the molecular topology belongs to an alternant or non-alternant hydrocarbon or which subset of carbon atoms in relation to the "starring" scheme the electrodes are connected to. 

In principle, it would also be possible to account for the presence of hetero atoms within the scheme as it has been done elsewhere~\cite{graphical3,graphical6} for a treatment of DTE containing its sulfur sites but this would come at the price of diminishing the scheme's simplicity and would not provide any important arguments for the present discussion. 

This AO scheme considers all orbitals in an AO representation and relies on the structure of the entire Hamiltonian thereby strongly reflecting the respective molecular topology. This is in contrast to any frontier orbital approximation within a MO based scheme where by definition all but two MOs are disregarded. The pairing theorem justifies this omission for the specific case, where each CR pair defined by the respective equal distance of its parts to $E_F$ cancels out individually for symmetry reasons. For the other cases where interference is constructive within each pair DQI can still occur due to cancellation between pairs. This is why all MO contributions are significant in this latter case as we will further explore below. 

\subsection{Convergence with respect to the number of included CR pairs of MOs}
\begin{figure*}
    \includegraphics[width=0.7\linewidth]{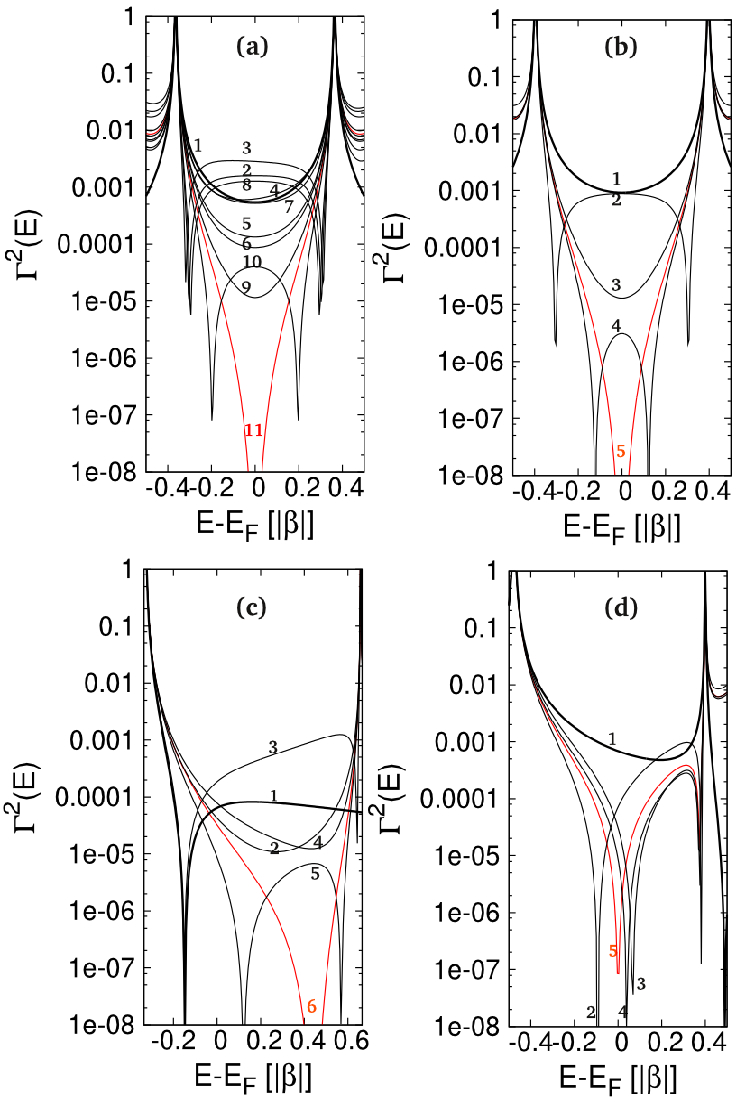}
    \caption{$\Gamma^2(E)$ (in arbitrary units) with an increasing number of CR pairs included in Eq.~\ref{gamma} for a) DNE in its open form, DTE in its open form: b) without S, c) with S, and d) azulene contacted in ($5$,$7$-) positions. The label $1$ means that only the HOMO and the LUMO enter Eq.~\ref{gamma}, for $2$ it is the HOMO, the LUMO, the HOMO-1 and the LUMO+1 and so on for higher labels up to where all CR pairs are included for the respective highest label. Only the lines with all CR pairs accounted for are shown in red, all other lines are in black.} 
    \label{fig5}
\end{figure*}

Independently of the frontier orbital approximation as we discuss it in this article, there is the common conception in studies of the conductance of single molecule junctions that $T(E_F)$ is dominated by the MOs close to $E_F$ since the tails of the peaks further apart decay rapidly and their contributions can therefore safely be disregarded~\cite{no2bipy,troels}. 

This assumption is also motivated by the fact that the respective distance of each MO to the Fermi energy enters its respective term in Eqns.~\ref{spectral} and~\ref{gamma} explicitly in the denominator and its increase can therefore be expected to reduce the terms significance. 

There are two reasons why such preconceptions should be questioned regarding their validity in general: i) While it is true that the denominator of a term in Eqns.~\ref{spectral} and~\ref{gamma} increases for high values of $\epsilon_{m}$, this effect might be outweighed by large couplings or large MO amplitudes at the contact sites; ii) the distinction between the occurrence and absence of DQI is often one between an exact value of zero (at least in the framework of TB where only $\pi$ electrons are considered) and a rather small number which appears to be bigger than it actually is due to the logarithmic plotting of $T(E)$. 

In Fig.~\ref{fig5} we increase the number of CR pairs of MOs included in the sum of Eq.~\ref{gamma} for the calculation of $\Gamma^2(E)$ stepwise for the systems in Fig.~\ref{fig2} which exhibit DQI close to the Fermi level but where this cannot be predicted within a frontier orbital approximation. Here we first consider only the HOMO and the LUMO (label $1$), then the HOMO, the LUMO, the HOMO-1 and the LUMO+1 (label $2$) and so on where only the red curve with the highest label includes the contributions from all CR pairs of MOs corresponding to the respective molecular topology. 

Quite contrarily to what might be expected, for the open forms of DNE (Fig.~\ref{fig5}a) and DTE (Fig.~\ref{fig5}b and c) the convergence of $\Gamma^2(E)$ with the number of included pairs is not smooth but oscillating because the contributions of CR pairs of MOs enter with alternating signs. Even for the label with only the CR pair with the energies most remote from $E_F$ missing, i.e. the labels $10$, $4$ and $5$ for DNE, DTE without and with S, respectively, the conductance is still far from zero on a logarithmic scale. For the label $3$ with the three CR pairs of MOs closest to the Fermi energy included it even has a magnitude comparable to that of the conducting closed form of the respective switch, where for DTE the convergence behavior seems to be rather unaffected by the presence or absence of the S atoms. 

This analysis strongly indicates that in order to capture DQI effects for a particular molecular topology correctly really all MOs belonging to its $\pi$ system need to be properly accounted for in order to achieve a reliable theoretical description. Even for the non-alternant hydrocarbon azulene contacted in ($5$,$7$-) positions (Fig.~\ref{fig5}d), where no destructive but also no constructive interference within each pair can be indicated directly from the pairing theorem, the contributions from the frontier orbitals, i.e. the HOMO and LUMO, alone (label $1$) do not result in any DQI feature close to $E_F$. The inclusion of the second CR pair of MOs (label $2$) produces this feature but it then again needs the contributions from all CR pairs to position it energetically exactly at $E_F$.

\begin{figure}
    \includegraphics[width=\linewidth]{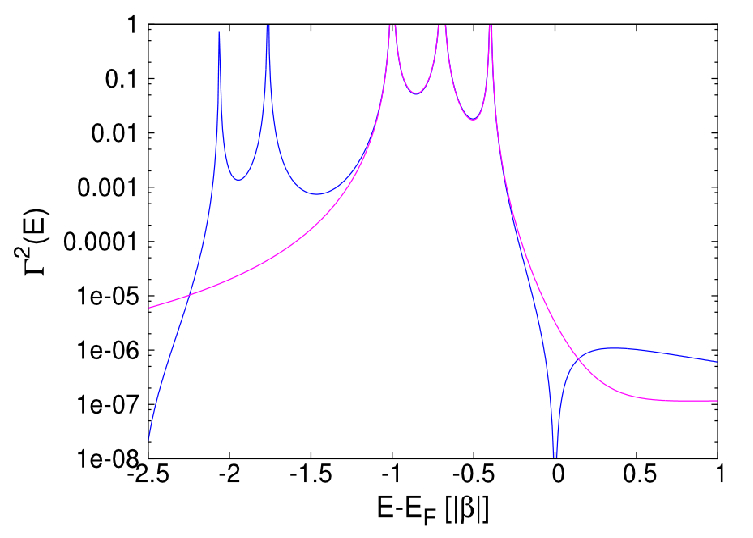}
    \caption{$\Gamma^2(E)$, where $\Gamma(E)$ is taken as the sum over all five occupied (blue curve) or only the three occupied MOs closest to $E_F$ (magenta curve) in Eq.~\ref{gamma} for the open form of the AH analog of DTE with the S atoms removed.}
    \label{fig6}
\end{figure}

Another property that arises from the pairing theorem is that in the case of DQI for even-membered AHs connected at sites belonging to different subsets, the contributions from all occupied and all unoccupied MOs to $\Gamma(E)$ in Eq.~\ref{gamma} must each cancel out individually at $E_F$. This is because in those cases the contribution from each half of a CR pair is equal to the other half in both sign and magnitude at $E_F$, which means that it is then sufficient to consider either all occupied or all unoccupied MOs alone. 

Making use of this knowledge, in Fig.~\ref{fig6} we plot $\Gamma^2(E)$ from the sum over the occupied states in Eqn.~\ref{gamma}, where we compare taking all five occupied MOs for the open form of the AH analog of DTE with the S atoms removed (blue curve) with the case where the lowest lying two MOs have been excluded from the summation (magenta curve). As can be seen from the figure, cutting out the lowest lying two MOs does not make any difference in the energy regions of the peaks of the three higher lying MOs, because the transmission in the vicinity of a peak is always largely dominated by the contribution of the one MO it is related to, but crucially decides whether DQI occurs at the Fermi level or not. 

This enforces the main message of our article that in order to identify DQI reliably from a MO perspective the contribution of all MOs need to be taken into account and not just a selected few of them. This finding has also high importance for the analysis of NEGF-DFT results, where a cut coupling approach is routinely used to describe DQI in terms of a few MOs close to $E_F$ only~\cite{no2bipy,troels}.

\section{Conclusions}

In summary, we showed that DQI in the electron transport of single molecule junctions can be reliably discussed from a MO perspective if the contributions from all MOs are accounted for and not only those from MOs close to the Fermi level. This applies in general and not only for even-membered alternant hydrocarbons contacted at carbon sites of the same subset as it is the case for predictions within a frontier orbital approximation. 

This MO perspective, however, does not in general allow for the prediction of DQI without prior numerical calculations within a TB framework which puts it into contrast to a recently proposed graphical AO scheme~\cite{graphical1,graphical2} where such predictions are indeed possible. On the other hand such a MO based analysis is not limited to the prediction of the zero-bias conductance defined by the transmission at the Fermi level and can thus provide essential input for the interpretation of computationally more demanding NEGF-DFT results as well as reconcile findings from single molecule electronics with more traditional concepts from quantum chemistry.

%\vspace{10 mm}

\begin{acknowledgments}
%\vspace{-10 mm}
X.Z. and R.S. are currently supported by the Austrian Science Fund FWF, project Nr. P27272. 
\end{acknowledgments}

%%%%%%% References

\bibliographystyle{apsrev}

\begin{thebibliography}{10}
\bibitem{header}S. Shaik, H. S. Rzepa and R. Hoffmann, Angew. Chem. Int. Ed. \textbf{52}, 3020 (2013).
\bibitem{chavy}C. Joachim, J. K. Gimzewski, R. R. Schittler, and C. Chavy, Phys. Rev. Letters {\bf 74}, 2102 (1995).
\bibitem{tour}M. A. Reed, C. Zhou, C. J. Muller, T. P. Burgin, and J. M. Tour, Science {\bf 278}, 252 (1997).
\bibitem{lohneysen}J. Reichert, R. Ochs, D. Beckman, H. B. Weber, M. Mayor, and H. v. Lohneysen, Phys. Rev. Letters {\bf 88}, 17680 (2002).
\bibitem{ruitenbeek}R. H. M. Smit, Y. Noat, C. Untiedt, N. D. Lang, M. C. van Hemert, and J. M. van Ruitenbeek, Nature {\bf 419}, 906 (2002).
\bibitem{keldysh}Y. Meir and N. S. Wingreen, Phys. Rev. Lett. \textbf{68}, 2512 (1992).
\bibitem{atk}M. Brandbyge, J. L. Mozos, P. Ordejon, J. Taylor and K.  Stokbro, Phys. Rev. B \textbf{65}, 165401 (2002).
\bibitem{xue}Y. Xue, S. Datta and M. A. Ratner, Chem. Phys. \textbf{281}, 151 (2002).
\bibitem{sanvito}A. R. Rocha, V. M. Garcia-Suarez, S. W. Baily, C. J. Lambert, J. Ferrer and S. Sanvito, Nature Materials \textbf{4}, 335 (2005).
\bibitem{kristian}K. S. Thygesen and K. W. Jacobsen, Chem. Phys. \textbf{319}, 111 (2005).
\bibitem{linderberg}Y. \"{O}hrn and J. Linderberg, Phys. Rev. 139, A1063 (1965).
\bibitem{joachim1}P. Sautet and C. Joachim, Chem. Phys. Lett. \textbf{153}, 511 (1988).
\bibitem{joachim2}C. Patoux, C. Coudret, J. P. Launay, C. Joachim and A. Gourdon, Inorg. Chem. \textbf{36}, 5037 (1997).
\bibitem{mayor}M. Mayor, H. B. Weber, J. Reichert, M. Elbing, C. von H\"anisch, D. Beckmann, and M. Fischer, Angew. Chem. Int. Ed. \textbf{42}, 5834 (2003).
\bibitem{coulson1}C. A. Coulson and H. C. Longuet-Higgins, Proc. R. Soc. Lond. A \textbf{191}, 39 (1947).
\bibitem{coulson2}C. A. Coulson and H. C. Longuet-Higgins, Proc. R. Soc. Lond. A \textbf{192}, 16 (1947).
\bibitem{coulson34}C. A. Coulson and H. C. Longuet-Higgins, Proc. R. Soc. Lond. A \textbf{193}, 447 (1948).
\bibitem{coulson5}C. A. Coulson and H. C. Longuet-Higgins, Proc. R. Soc. Lond. A \textbf{195}, 188 (1948).
\bibitem{coulson6}B. H. Chirgwin and C. A. Coulson, Proc. R. Soc. Lond. A \textbf{201}, 196 (1950).
\bibitem{pickup}B. T. Pickup, Philos. Mag. B \textbf{69}, 799 (1994).
\bibitem{stuyver1}T. Stuyver, S. Fias, F. De Proft, P. W. Fowler and P. Geerlings, J. Chem. Phys. \textbf{142}, 094103 (2015).
\bibitem{stuyver2}T. Stuyver, S. Fias, F. De Proft and P. Geerlings, Chem. Phys. Lett. \textbf{630}, 51 (2015).
\bibitem{stuyver3}T. Stuyver, S. Fias, F. De Proft and P. Geerlings, J. Phys. Chem. C \textbf{119}, 26390 (2015).
\bibitem{diradical}Y. Tsuji, R. Hoffmann, M. Strange and G. C. Solomon, Proc. Natl. Acad. Sci. USA \textbf{113}, E413 (2016).
\bibitem{datta}S. Datta, {\it Electronic transport in mesoscopic systems}, (Cambridge Univ. Press, Cambridge, UK, 1995).
\bibitem{lambert}D. Z. Manrique, C. Huang, M. Baghernejad, X. Zhao, O. A. Al-Owaedi, H. Sadeghi, V. Kaliginedi, W. Hong, M. Gulcur, T. Wandlowski, M. R. Bryce, C. J. Lambert, Nat. Commun. \textbf{6}, 6389 (2015).
\bibitem{reuter}M. G. Reuter and T. Hansen, J. Chem. Phys. \textbf{141}, 181103 (2014).
\bibitem{memory}R. Stadler, M. Forshaw and C. Joachim, Nanotechnology \textbf{14}, 138 (2003).
\bibitem{fano}R. Stadler and T. Markussen, J. Chem. Phys. \textbf{135}, 154109 (2011).
\bibitem{graphical1}R. Stadler, S. Ami, M. Forshaw and C. Joachim, Nanotechnology \textbf{15}, S115 (2004).
\bibitem{graphical2}T. Markussen, R. Stadler and K. S. Thygesen, Nano Lett. \textbf{10}, 4260 (2010).
\bibitem{graphical3}T. Markussen, R. Stadler and K. S. Thygesen, Phys. Chem. Chem. Phys. \textbf{13}, 14311 (2011).
\bibitem{graphical4}R. Stadler, Nano Lett. \textbf{15}, 7175 (2015).
\bibitem{graphical5}K. G. L. Pedersen, A. Borges, P. Hedeg\r{a}rd, G. C. Solomon and M. Strange, J. Phys. Chem. C \textbf{119}, 26919 (2015).
\bibitem{yoshizawa1}K. Yoshizawa, T. Tada and A. Staykov, J. Am. Chem. Soc. \textbf{130}, 9406 (2008).
\bibitem{yoshizawa2}Y. Tsuji, A. Staykov and K. Yoshizawa, Thin solid films \textbf{518}, 444 (2009).
\bibitem{yoshizawa3}Y. Tsuji, A. Staykov and K. Yoshizawa, J. Phys. Chem. C \textbf{113}, 21477 (2009).
\bibitem{yoshizawa4}X. Li, A. Staykov and K. Yoshizawa, J. Phys. Chem. C \textbf{114}, 9997 (2010).
\bibitem{yoshizawa5}Y. Tsuji, A. Staykov and K. Yoshizawa, J. Am. Chem. Soc. \textbf{133}, 5955 (2011).
\bibitem{yoshizawa6}K. Yoshizawa, Acc. Chem. Res. \textbf{45}, 1612 (2012).
\bibitem{vanvleck}J. H. Van Vleck and A. Sherman, Rev. Mod. Phys. {\textbf 7}, 167 (1935).
\bibitem{pairing}C. A. Coulson and G. S. Rushbrooke, Math. Proc. Cambridge Philos. Soc. \textbf{36}, 193 (1940).
\bibitem{proof}I. Gutman and O. E. Polansky, {\it Mathematical concepts in organic chemistry}, p. 57-58 (Springer, 1986), ISBN: 978-3-642-70984-5.
\bibitem{larsson}S. Larsson, J. Am. Chem. Soc. \textbf{103}, 4034 (1981).
\bibitem{ratner}M. A. Ratner, J. Phys. Chem. \textbf{94}, 4877 (1990).
\bibitem{georg}G. Kastlunger and R. Stadler, Phys. Rev. B \textbf{89}, 115412 (2014).
\bibitem{sautet1}P. Sautet and M.-L. Bocquet, Phys. Rev. B \textbf{53}, 4910 (1996).
\bibitem{no2bipy}R. Stadler, Phys. Rev. B 80, 125401 (2009).
\bibitem{graphical6}M. H. Garner, G. C. Solomon and M. Strange, J. Phys. Chem. C \textbf{120}, 9097 (2016).
\bibitem{fermi}R. Stadler, J. Phys.: Conf. Ser. \textbf{61}, 1097 (2007). 
\bibitem{mikkel1}K. G. L. Pedersen, M. Strange, M. C. Leijnse, P. Hedeg\r{a}rd, G. C. Solomon and J. Paaske, Phys. Rev. B \textbf{90}, 125413 (2014).
\bibitem{mikkel2}M. Strange, J. S. Seldenthuis, C. J. O. Verzijl, J. M. Thijssen and G. C. Solomon, J. Chem. Phys. \textbf{142}, 084703 (2015).
\bibitem{polyenes}Y. Tsuji, R. Hoffmann, R. Movassagh and S. Datta, J. Chem. Phys. \textbf{141}, 224311 (2014).
\bibitem{sautet2}P. Sautet and C. Joachim, Phys. Rev. B \textbf{38}, 12238 (1988).
\bibitem{ase1}S. R. Bahn and K. W. Jacobsen, Comput. Sci. Eng. \textbf{4}, 56 (2002).
\bibitem{ase2}https://wiki.fysik.dtu.dk/gpaw/exercises/transport/
transport.html
\bibitem{hulis}http://www.hulis.free.fr/
\bibitem{switches}S. J. van der Molen and P. Liljeroth, J. Phys.: Cond. Mat. \textbf{22}, 133001 (2010).
\bibitem{irie}M. Irie, Chem. Rev. \textbf{100}, 1685 (2000).
\bibitem{irie1}K. Matsuda and M. Irie, J. Am. Chem. Soc. \textbf{122}, 7195 (2000).
\bibitem{launay}S. Fraysse, C. Coudret and J.P. Launay, Eur. J. Inorg. Chem., 1581 (2000)
\bibitem{dulic}D. Dulić, S. J. van der Molen, T. Kudernac, H. T. Jonkman, J. J. D. de Jong, T. N. Bowden, J. van Esch, B. L. Feringa, and B. J. van Wees, Phys. Rev. Lett. \textbf{91}, 207402 (2003).
\bibitem{victor}C. Van Dyck, V. Geskin, A. J. Kronemeijer, D. M. de Leeuw and J. Cornil, Phys. Chem. Chem. Phys. \textbf{15}, 4392 (2013).
\bibitem{angewandte1}Y. Tsuji and R. Hoffmann, Angew. Chem. Int. Ed. \textbf{53}, 4093 (2014).
\bibitem{azulene}J. Xia, B. Capozzi, S. Wei, M. Strange, A. Batra, J. R. Moreno, R. J. Amir, E. Amir, G. C. Solomon, L. Venkataraman and L. M. Campos, Nano. Lett. \textbf{14}, 2941 (2014).
\bibitem{angewandte2}F. Schwarz, M. Koch, G. Kastlunger, K. Venkatesan, H. Berke, R. Stadler and E. L\"{o}rtscher, Angew. Chem. Int. Ed., in print (2016); DOI: 10.1002/anie.201605559
\bibitem{troels}C. M. Guédon, H. Valkenier, T. Markussen, K. S. Thygesen, J. C. Hummelen and S. J. van der Molen, Nat. Nanotechnol. \textbf{7}, 305 (2012).
\end{thebibliography}

\expandafter\ifx\csname natexlab\endcsname\relax\global\long\def\natexlab#1{#1}
 \fi \expandafter\ifx\csname bibnamefont\endcsname\relax \global\long\def\bibnamefont#1{#1}
 \fi \expandafter\ifx\csname bibfnamefont\endcsname\relax \global\long\def\bibfnamefont#1{#1}
 \fi \expandafter\ifx\csname citenamefont\endcsname\relax \global\long\def\citenamefont#1{#1}
 \fi \expandafter\ifx\csname url\endcsname\relax \global\long\def\url#1{\texttt{#1}}
 \fi \expandafter\ifx\csname urlprefix\endcsname\relax\global\long\def\urlprefix{URL }
 \fi \providecommand{\bibinfo}[2]{#2} \providecommand{\eprint}[2][]{\url{#2}} 

\end{document}